\documentclass{article}

\usepackage{arxiv}

\usepackage[utf8]{inputenc} 
\usepackage[T1]{fontenc}    
\usepackage{hyperref}       
\usepackage{url}            
\usepackage{booktabs}       
\usepackage{amsfonts}       
\usepackage{nicefrac}       
\usepackage{microtype}      
\usepackage{lipsum}		
\usepackage{graphicx}
\usepackage{natbib}
\usepackage{doi}
\usepackage{physics}
\usepackage{float}
\usepackage{subcaption}
\usepackage{comment}
\usepackage{multirow}
\usepackage{xcolor}
\newcommand{\argmin}{\mathop{\mathrm{argmin}}}

\title{Physics-Informed Neural Networks with\\ Adaptive Localized Artificial Viscosity}

\author{
	Emilio Jose Rocha Coutinho\\
	Department of Petroleum Engineering\\
	Texas A\&M University and Petrobras\\
	emiliocoutinho@gmail.com\\
	\\
	\And
	Marcelo Dall'Aqua\\
	Department of Petroleum Engineering\\
	Texas A\&M University\\
	marcelo.dallaqua@tamu.edu\\
	\\
	\And
	Levi McClenny\\
	Department of Electrical \& Computer Engineering\\
	Texas A\&M University\\
    levimcclenny@tamu.edu\\
    \\
    \And
    Ming Zhong\\
    Texas A\&M Institute of Data Science\\
	Texas A\&M University\\
	mingzhong@tamu.edu\\
	\\
	\And
	Ulisses Braga-Neto\\
	Department of Electrical \& Computer Engineering\\
	Texas A\&M University\\
	ulisses@tamu.edu\\
	\\
	\And
	Eduardo Gildin\\
	Department of Petroleum Engineering\\
	Texas A\&M University\\
	egildin@tamu.edu\\
	\\
}

\date{}

\renewcommand{\shorttitle}{PINN with Adaptive Localized Artificial Viscosity}

\hypersetup{
    pdftitle={Physics-Informed Neural Networks with Adaptive Localized Artificial Viscosity},
    pdfsubject={},
    pdfauthor={Emilio J. R. Coutinho, Marcelo Dall'Aqua, Levi McClenny, Ming Zhong, Ulisses Braga-Neto, Eduardo Gildin},
    pdfkeywords={Physics-Informed Neural Networks, Artificial Viscosity, Hyperbolic PDEs},
}

\begin{document}
\maketitle

\begin{abstract}
	Physics-informed Neural Network (PINN) is a promising tool that has been applied in a variety of physical phenomena described by partial differential equations (PDE). However, it has been observed that PINNs are difficult to train in certain ``stiff'' problems, which include various nonlinear hyperbolic PDEs that display shocks in their solutions. Recent studies added a diffusion term to the PDE, and an artificial viscosity (AV) value was manually tuned to allow PINNs to solve these problems. In this paper, we propose three approaches to address this problem, none of which rely on an a priori definition of the artificial viscosity value. The first method learns a global AV value, whereas the other two learn localized AV values around the shocks, by means of a parametrized AV map or a residual-based AV map. We applied the proposed methods to the inviscid Burgers equation and the Buckley-Leverett equation, the latter being a classical problem in Petroleum Engineering. The results show that the proposed methods are able to learn both a small AV value and the accurate shock location and improve the approximation error over a nonadaptive global AV alternative method. 
\end{abstract}

\keywords{Physics-Informed Neural Networks \and Artificial Viscosity \and Hyperbolic PDEs}

\section{Introduction}
\label{sec:intro}

Over the past ten years, we have seen a substantial increase in the use of machine learning methods in science and engineering areas. However, most of these methods are data-driven, which can lead to unrealistic or non-physical models. Scientists and engineers use mathematical tools to model physical phenomena to make predictions and support decisions. Modeling the physical systems, understanding how the models behave, how they can be solved accurately and in a reasonable time, and how uncertainty should be considered are some of the areas that have been developed in the last decades. A significant challenge in applying machine learning methods to scientific and engineering problems is to honor physical knowledge. Machine learning techniques can provide fast and accurate results compared to traditional scientific computation methods. However, if these results do not align with the understanding of physical phenomena, they will not be of much use.  

Physics-Informed Neural Networks (PINNs) use knowledge about the physical phenomena in building a machine learning model. PINNs can model phenomena described by partial differential equations with boundary and initial conditions by incorporating these constraints in the loss function of an artificial neural network. A PINN can integrate sensor data, if available, or it can be employed as a traditional PDE solver without any data beyond the initial and boundary conditions. In this paper, we focus on PINNs as a PDE solver. Like any other scientific computation method, it has strengths and limitations. Its greatest strength as a PDE solver is the ability to model high-dimensional problems with complicated boundaries since a PINN is a meshless method that does not require the construction of elaborate grids. A significant limitation of PINNs is the difficulty of training the neural network in the presence of sharp transitions in the solution, such as shocks and contact discontinuities in nonlinear hyperbolic conservation laws.

We propose three methods to address this limitation, which learn the necessary artificial viscosity to be applied to the PDE during the PINN training procedure. The first method learns a global AV value, while the other two can localize the artificial diffusion to the areas in the solution domain where its presence is necessary. The amount and location of the added artificial viscosity are controlled by an adaptive viscosity coefficient, which is learned automatically during the neural network training procedure, producing an artificial viscosity map. We report the results of applying the methods to two classical hyperbolic PDEs involving shocks, namely, the inviscid Burgers equation and the Buckley-Leverett equation. The latter is a well-known problem in the Petroleum industry. 
The results show that the proposed methods are able to learn both a small AV value and the accurate shock location and improve the approximation error over a nonadaptive global AV alternative method.

\section{Related work}
\label{sec:rel_work}

Taking advantage of Artificial Neural Networks as a universal function approximators \citep{hornik_multilayer_1989}, Physics-informed Neural Networks (PINN) \citep{raissi_physics-informed_2019} were proposed to solve complex physical problems that are modeled using Partial Differential Equations along with their initial and boundary conditions. PINNs have been used in a variety of scientific and engineering problems, such as heat transfer \citep{cai_physics-informed_2021} fluid flow in porous media \citep{almajid_prediction_2020}, and weather and climate modeling \citep{kashinath_physics-informed_2021}. Limitations of the method have mainly to do with the difficulty of successfully training the neural network. Various approaches to address this problem have been proposed \citep{wang_understanding_2021,liu_dual-dimer_2021,wang_when_2022,mcclenny_self-adaptive_2020,davi_pso-pinn_2022}.

Regarding hyperbolic PDEs, studies \citep{fuks_limitations_2020, fraces_physics_2020, fraces_physics_2021} have been conducted to employ PINNs to solve a classic problem in petroleum reservoir engineering called the Buckley-Leverett equation \citep{buckley_mechanism_1942}. These studies have shown that PINNs fail to find the solution of the PDE when it has hyperbolic behavior with shocks and contact discontinuities in the solution. To address this issue, they added a diffusion term to the PDE. While the method produces good results, it is highly dependent on an apriori choice of the artificial diffusion coefficient \citep{patel_thermodynamically_2020}. Another disadvantage of the method in \citep{fuks_limitations_2020, fraces_physics_2020, fraces_physics_2021} is the application of artificial diffusion to the entire solution domain, rather than only near the jumps in the solution, as recommended, for example, in \citep{reisner_spacetime_2013}. Adding artificial viscosity to reduce the hyperbolicity of the PDE is a well-known approach in traditional scientific computation \citep{reisner_spacetime_2013, harlow_numerical_1971, gentry_eulerian_1966, stiernstrom_residual-based_2021}. In the context of PINNs, \cite{fraces_physics_2021} proposed using Welge's method \citep{welge_simplified_1952} to handle the shock front in the Buckley-Leverett problem. Welge's method transforms the fractional flow function to assure that the entropy condition is satisfied; this method is only valid with homogeneous initial conditions. \cite{rodriguez-torrado_physics-informed_2021} proposed another approach to solving the problem by enforcing initial and boundary conditions and removing them from a recurrent neural network's training procedure, which had only the residual term on its loss function. This approach seems to violate the PINNs original proposition and may lead to severe inconsistencies when obtaining estimation close to the initial and boundaries. The method accuracy depends on the resolution of the domain discrete version, like the numerical methods depend on the domain discretization, which is a critical issue in solving hyperbolic problems using traditional numerical methods.

\section{Background}
\label{sec:background}

This section describes hyperbolic conservation laws, which is the paper's focus, followed by a brief review of Physics-informed Neural networks (PINN).

\subsection{Hyperbolic Conservation Law PDEs}

A general hyperbolic conservation law PDE reads as follows:
\begin{align}
    &\pdv{u(\vb*{x},t)}{t} + \pdv{f(u(\vb*{x},t))}{x} = 0, \label{eq:pde}
\end{align}
where $u$ is the quantity we would like to solve for, $\vb*{x}$ is the space variable, $t$ is the time variable, and $f(u(\vb*{x},t))$ is a flux function. Additionally, there may be initial and boundary conditions, which are problem-dependent. The solution of hyperbolic PDEs can sometimes be obtained analytically by the method of characteristics \citep{lax_hyperbolic_1973}.

The solution of nonlinear hyperbolic PDEs can develop shocks and contact discontinuities, making them very challenging to numerical methods. One way to address this difficulty is to reduce the hyperbolicity of the problem by adding to the PDE an artificial  diffusion term modulated by a small viscosity coefficient $\nu>0$: 
\begin{align}
    &\pdv{u(\vb*{x},t)}{t} + \pdv{f(u(\vb*{x},t))}{\vb*{x}} = \nu \pdv[2]{u(\vb*{x},t)}{\vb*{x}}\,. \label{eq:pde_parabolic}
\end{align}

By virtue of adding the diffusion term, it can be shown that the system in (\ref{eq:pde_parabolic}) has a solution that is free of discontinuities. However, the artificial viscosity is applied to the whole solution domain in (\ref{eq:pde_parabolic}). The nonlocalized application of the artificial viscosity can lead to errors or undesired smoothness. Consequently, it should be used only in regions where jumps in the solution due to shocks and contact discontinuities occur~\citep{reisner_spacetime_2013}.

\subsection{Physics-informed Neural Network}

PINNs can be applied to approximate solutions of the PDEs discussed in the previous section. Consider a general PDE with initial and boundary conditions:
\begin{subequations}
    \label{eq:pinn_eq_system}
    \begin{align}
    &\pdv{u(\vb*{x},t)}{t} + \mathcal{N}_{\vb*{x}}(u(\vb*{x},t)) = 0, \quad \vb*{x} \in \Omega, t \in [0, T] \label{eq:pinn_pde}\\
    &u(\vb*{x}, t) = g(\vb*{x}, t), \quad \vb*{x} \in \partial \Omega, t \in [0, T] \label{eq:pinn_pde_cc}\\
    &u(\vb*{x}, 0) = u_0(\vb*{x}) , \quad \vb*{x} \in \Omega \label{eq:pinn_pde_ci}
\end{align}
\end{subequations}
where $u({\vb*{x},t})$ is the unknown, $\Omega$ is the domain of definition of the problem, $\vb*{x}$ is a spatial vector variable, $t$ is time, and $\mathcal{N}_{\vb*{x}}(\cdot)$ is a differential operator. Equations \ref{eq:pinn_pde_cc} and \ref{eq:pinn_pde_ci} provide boundary and initial conditions, respectively. We would like to approximate $u(\vb*{x},t)$ by the output of a {\em physics-informed neural network} $u(\vb*{x},t, \vb*{w})$ with parameters $\vb*{w}$ (weights and biases) \citep{raissi_physics-informed_2019}. 

A traditional machine learning application requires training data, which could be data obtained from the analytical or numerical solution of the PDE. Using this data, one would train the NN to predict $u(\vb*{x},t)$. The loss function used in the traditional neural network approach is
\begin{equation}
    \mathcal{L}_{data}(\vb*{w}) = \frac{1}{N}\sum_{i=1}^{N}{\left(u(\vb*{x}_s^i, t_s^i, \vb*{w})-y_s^i\right)^2}, \label{eq:nn_loss}
\end{equation}
where $\{(\vb*{x}_s^i,t_s^i,y_s^i)\}_{i=1}^{N_s}$ is the training data. However, this approach can fail in complex applications due to the large amount of data needed to capture the system dynamics.
The PINN approach, on the other hand, has a much smaller data requirement, or indeed no data requirement, in case the physics of the problem is completely known and expressed in a PDE system such as the one in (\ref{eq:pinn_eq_system}). A PINN does this by incorporating the physics embedded in the PDE, boundary, and initial conditions into the loss function: 
\begin{equation}
    \mathcal{L}(\vb*{w}) = \mathcal{L}_r(\vb*{w}) + \mathcal{L}_b(\vb*{w}) + \mathcal{L}_0(\vb*{w}) , \label{eq:pinn_loss_total}
\end{equation}
where $\mathcal{L}_r(\vb*{w})$ is the loss corresponding to the residual of the PDE, $\mathcal{L}_b(\vb*{w})$ is the loss due to boundary conditions and $\mathcal{L}_0(\vb*{w})$ is the loss function due to the initial condition, given by:
\begin{equation}
    \mathcal{L}_r(\vb*{w}) = \frac{1}{N_r}\sum_{i=1}^{N_r}{r(\vb*{x}_r^i, t_r^i, \vb*{w})^2},  \label{eq:pinn_loss_residual}
\end{equation}
\begin{equation}
    \mathcal{L}_b(\vb*{w}) = \frac{1}{N_b}\sum_{i=1}^{N_b}{\left(u(\vb*{x}_b^i, t_b^i, \vb*{w})-g(\vb*{x}_b^i, t_b^i)\right)^2}, \label{eq:pinn_loss_boundary}
\end{equation}
\begin{equation}
    \mathcal{L}_0(\vb*{w}) = \frac{1}{N_0}\sum_{i=1}^{N_0}{\left(u(\vb*{x}_0^i, 0, \vb*{w})-u_0(\vb*{x}_0^i)\right)^2}, \label{eq:pinn_loss_initial}
\end{equation}
where $\{\vb*{x}_r^i, t_r^i\}_{i=1}^{N_r}$ is a set of collocation points where the PDE is enforced, $\{\vb*{x}_b^i, t_b^i\}_{i=1}^{N_b}$ is a set of points to enforce the boundary conditions, and  $\{\vb*{x}_0^i\}_{i=1}^{N_0}$ is a set of points in the domain to enforce the initial condition. These points are randomly selected in the residual, boundary and initial conditions domains. The derivatives of the network output needed to obtain the PDE residue are computed using automatic differentiation.

\section{Methodology}
\label{sec:methodology}

This section presents our contributions to the solution of hyperbolic PDEs using PINNs. First, we introduce the idea of adaptive artificial viscosity that can be learned during the training procedure and does not depend on an apriori choice of artificial viscosity coefficient. We then show how to localize the artificial viscosity only in regions close to discontinuities. Based on these ideas, we propose three methods for training PINNs with a learnable and/or localized artificial viscosity. These methods take advantage of the power of the optimization algorithm used in deep neural network training to learn the necessary artificial viscosity. 

\subsection{PINN with Learnable Global Artificial Viscosity}
\label{sec:pinn_learnable_av}

In this case, a single value for the artificial viscosity coefficient $\nu$ is learned during the PINN training procedure. Following (\ref{eq:pde_parabolic}), the PDE residue loss is given by:
\begin{align}
    r(\vb*{x}, t, \vb*{w}, \nu) \,=\, 
    \pdv{u(\vb*{x},t, \vb*{w})}{t} + \pdv{f(u(\vb*{x},t,\vb*{w}))}{\vb*{x}} - \nu \pdv[2]{u(\vb*{x},t, \vb*{w})}{x}\,,
\end{align}
where the artificial viscosity coefficient $\nu$ is treated as an unknown parameter and determined by gradient descent, using the Adam algorithm, together with the neural network weights and biases. The training loss function is:
\begin{equation}
    \mathcal{L}(\vb*{w}, \nu) = \mathcal{L}_r(\vb*{w}, \nu) + \mathcal{L}_b(\vb*{w}) + \mathcal{L}_0(\vb*{w}) + \alpha_{visc} \mathcal{L}_{visc}(\nu), \label{eq:pinn_loss_local}
\end{equation}
where $\alpha_{visc}$ is a training weight applied to the artificial viscosity loss function, which here we set to $\mathcal{L}_{visc}(\nu)= \nu^2$. This loss component is a penalty term introduced to keep the value of the viscosity $\nu$ small (other choices of penalty are possible, such as the absolute value of $\nu$). The other loss components are as in (\ref{eq:pinn_loss_residual})--(\ref{eq:pinn_loss_initial}). The value of the penalty coefficient $\alpha_{visc}$ can be pre-specified or also be learned during the training procedure, as detailed later. 

\subsection{PINN with Parametric Artificial Viscosity Map}
\label{sec:pinn_param_map}

As mentioned previously, ideally the artificial viscosity should not be applied to the entire solution domain \citep{reisner_spacetime_2013}. Our first attempt to localize the application of artificial viscosity is to build a map that is parametrized by information from the problem solution. The PDE residual equation for this method can be written as:
\begin{align}
    r(\vb*{x}, t, \vb*{w}, \nu_{\max},\vb*{\theta}) \,=\, 
    \pdv{u(\vb*{x},t, \vb*{w})}{t} + \pdv{f(u(\vb*{x},t,\vb*{w}))}{\vb*{x}} - \nu_{\max}\nu(\vb*{x}, t,\vb*{\theta})\pdv[2]{u(\vb*{x},t, \vb*{w})}{x},
\label{eq:res_map}
\end{align}
where $\nu(\vb*{x},t,\vb*{\theta})$ is a spatial-temporal artificial viscosity map that has its values bounded between 0 and 1, $\nu_{\max}>0$ is the maximum allowed amount of artificial viscosity, and $\vb*{\theta}$ is a parameter vector. Both $\nu_{\max}$ and $\vb*{\theta}$ are treated as unknown parameters to be determined by gradient descent during neural network training. The training loss is defined similarly as in (\ref{eq:pinn_loss_local}):
\begin{equation}
    \mathcal{L}(\vb*{w}, \nu_{\max},\vb*{\theta}) \,=\, \mathcal{L}_r(\vb*{w}, \nu_{\max},\vb*{\theta}) + \mathcal{L}_b(\vb*{w}) + \mathcal{L}_0(\vb*{w}) + \alpha_{visc} \mathcal{L}_{visc}(\nu_{\max}), \label{eq:pinn_loss_map}
\end{equation}
with $\mathcal{L}_{visc}(\nu_{\max})= \nu_{\max}^2$, though, as before, other choices of penalty are possible. Note that when computing the residual loss component, the artificial viscosity map is only evaluated at the residue collocation points.

\begin{figure}[!ht]
    \centering
    \begin{subfigure}{0.4\textwidth}
        \includegraphics[width=\linewidth]{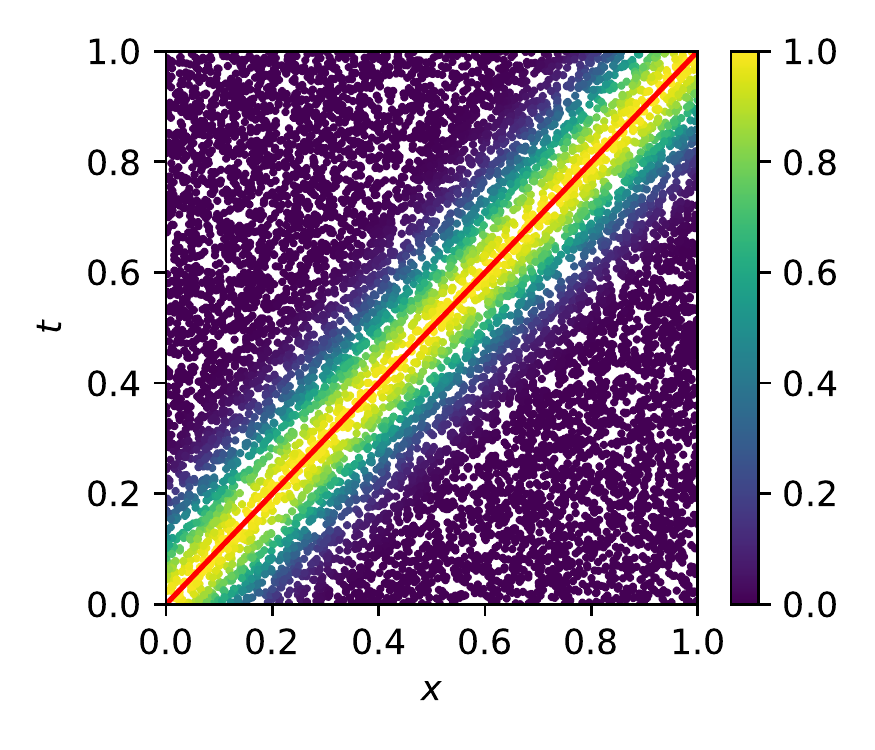}
        \caption{Artificial viscosity map}
        \label{fig:av_parametrized_map}
    \end{subfigure}
    \begin{subfigure}{0.4\textwidth}
        \includegraphics[width=\linewidth]{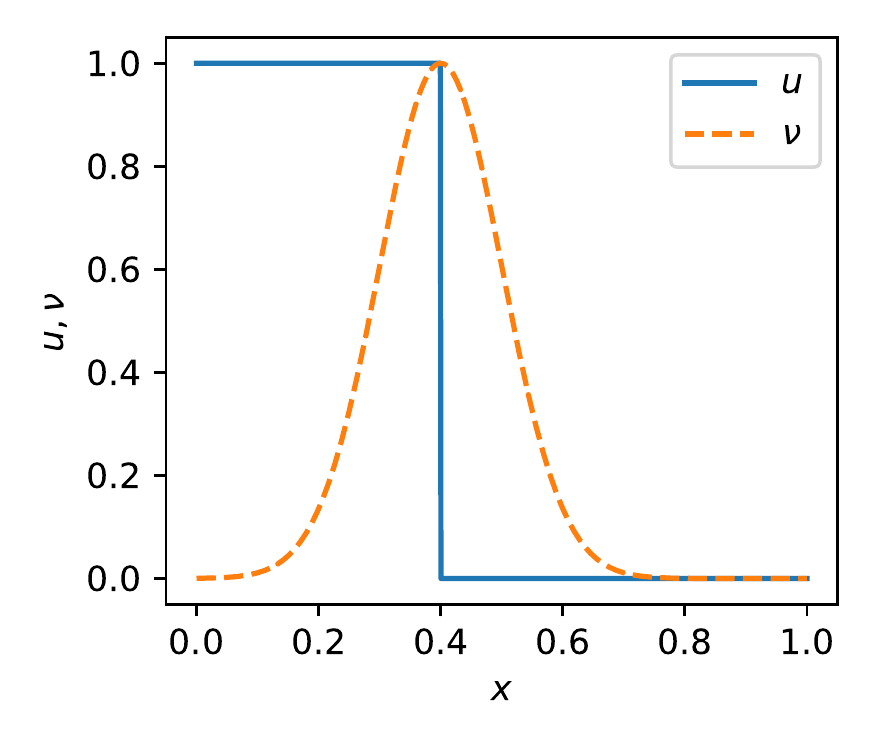}
        \caption{Artificial viscosity profile}
        \label{fig:av_parametrized_map_profile}
    \end{subfigure}
    \caption{(a) Artificial viscosity map built with parameters $v_{shock} = 1.0$, $w_\nu = 0.1$. The red diagonal line represents the shock front path of a hypothetical problem. (b) Artificial viscosity profile (orange dashed line) along $x$ direction at time $t=0.4$ built using parameters $w_\nu = 0.1$. The blue line represents the analytical solution of the hypothetical problem.}
    \label{fig:av_parametrized}
\end{figure}

To provide an example of a parametric viscosity map, we use the Buckley-Leverett problem (see Section \ref{sec:buckley_leverett} for more details about this problem). In this case, we know that if a shock front exists, it will form immediately after the initial time and will have a constant velocity. With this information, it is possible to determine the shock front path, which defines the region where the viscosity should be applied, while no artificial viscosity is applied to other areas of the solution. Let the shock front velocity $v_{shock}$ and the artificial viscosity bandwidth $w_\nu$ be the parameters needed to build the map. Figure \ref{fig:av_parametrized_map} displays an example of artificial viscosity map corresponding to $v_{shock} = 1.0$ and $w_\nu = 0.1$. The distribution of artificial viscosity values along the $x$ dimension at a specific time is based on a Gaussian probability density function centered on the shock front position with a standard deviation of $w_\nu$, as seen in Figure \ref{fig:av_parametrized_map_profile}, for time $t=0.4$.
In an actual problem, the shock front velocity $v_{shock}$ is not known. Therefore, it is learned as a parameter during neural network training. The same can be done for the artificial viscosity bandwidth $w_\nu$, though in the results reported in this paper, we set $w_\nu = 0.1$. 

\subsection{PINN with Residual-based Artificial Viscosity Map}
\label{sec:pinn_residual_map}

The previous method is effective, though it requires information about the structure of the solution, such as the number of shocks. As an alternative, we consider a nonparametric residual-based artificial viscosity method in this section. The use of the partial differential equation residue to localize the application of the artificial viscosity has been proven efficient in solving hyperbolic problems using numerical methods \citep{stiernstrom_residual-based_2021, bruno_fc-based_2021, nazarov_residual-based_2013, guo_adaptive_2017}. 

Here, the residue is similar to that in (\ref{eq:res_map}),
\begin{align}
    r(\vb*{x}, t, \vb*{w}, \nu_{\max}) \,=\, \pdv{u(\vb*{x},t, \vb*{w})}{t} + \pdv{f(u)}{x} - \nu_{\max}\nu(\vb*{x},t,\vb*{w})\pdv[2]{u(\vb*{x},t, \vb*{w})}{x} \label{eq:pinn_residual_pde_map},
\end{align}
 where $\nu_{\max}$ is again a learnable parameter that represents the maximum allowable value of artificial viscosity, while the training loss is given by
\begin{equation}
    \mathcal{L}(\vb*{w}, \nu_{\max}) \,=\, \mathcal{L}_r(\vb*{w}, \nu_{\max}) + \mathcal{L}_b(\vb*{w}) + \mathcal{L}_0(\vb*{w}) + \alpha_{visc} \mathcal{L}_{visc}(\nu_{\max})\,. \label{eq:pinn_loss_residualbased_av}
\end{equation}
Notice that this approach is nonparametric in the sense that there is no extra parameter vector $\vb*{\theta}$. Instead, the artificial viscosity map depends directly on the approximating neural network solution. Here, we adapt for PINN training the method proposed by \cite{stiernstrom_residual-based_2021} and set
\begin{equation}
    \nu(\vb*{x}, t,\vb*{w}) \,=\, \min{\left(\nu_1(\vb*{x}, t,\vb*{w}), \nu_r(\vb*{x}, t,\vb*{w})\right)} 
\label{eq:av_residual},    
\end{equation}
where $\nu_1(\vb*{x}, t,\vb*{w})$ is called first-order viscosity vector and $\nu_r(\vb*{x}, t,\vb*{w})$ is the high-order residual viscosity vector \citep{stiernstrom_residual-based_2021}. At each collocation point $(\vb*{x}_r^i, t_r^i)$ the first-order viscosity is calculated by:
\begin{equation}
    \nu_1(\vb*{x}_r^i, t_r^i,\vb*{w}) \,=\, \max_{loc}\,\{{\left|f'(u(\vb*{x}_r^i, t_r^i,\vb*{w}))\right|\}},    
\end{equation}
where the notation $\max_{loc}$ represents the maximum value taken over the neighbors of the collocation point $(\vb*{x}^i, t^i)$. This parameter is an estimate of the shock speed and serves as an upper bound for the artificial viscosity. The high-order residual viscosity at collocation point $(\vb*{x}_r^i, t_r^i)$ is defined as:
\begin{equation}
    \nu_r(\vb*{x}_r^i, t_r^i,\vb*{w}) \,=\, \max_{loc}\,\left\{\frac{\left|\bar{r}(\vb*{x}_r^i, t_r^i, \vb*{w})\right|}{n(\vb*{x}_r^i, t_r^i, \vb*{w})}\right\},
\end{equation}
where $\bar{r}(\vb*{x}_r^i, t_r^i, \vb*{w})$ is the inviscid PDE residual:
\begin{align}
    \bar{r}(\vb*{x}, t, \vb*{w}) \,=\, 
    \pdv{u(\vb*{x},t, \vb*{w})}{t} + \pdv{f(u(\vb*{x},t,\vb*{w}))}{\vb*{x}}\,,
\label{eq:res_inviscid}
\end{align}
and the normalization term $n(\vb*{x}_r^i, t_r^i, \vb*{w})$ is given by:
\begin{equation}
    n(\vb*{x}_r^i, t_r^i, \vb*{w}) \,=\, \left|\tilde{u}(\vb*{x}_r^i, t_r^i, \vb*{w}) - \max_{j=1,\ldots,N_r} \left\{\left|u(\vb*{x}_r^j, t_r^j, \vb*{w})-\frac{1}{N_r}\sum_{k=1}^{N_r}u(\vb*{x}_r^k, t_r^k, \vb*{w})\right| \right\}\right|, 
\end{equation}
where
\begin{equation}
    \tilde{u}(\vb*{x}_r^j, t_r^j, \vb*{w}) \,=\, \max_{loc}\,\{u(\vb*{x}_r^i, t_r^i, \vb*{w})\} - \min_{loc}\,\{u(\vb*{x}_r^i, t_r^i, \vb*{w})\}\,, 
\end{equation}
with $\min_{loc}$ being defined similarly as $\max_{loc}$. Finally, the artificial viscosity map obtained from (\ref{eq:av_residual}) is scaled linearly so that its minimum and maximum correspond to 0 and 1, respectively.

\begin{figure}[!ht]
    \centering
    \begin{subfigure}{0.4\textwidth}
        \includegraphics[width=\linewidth]{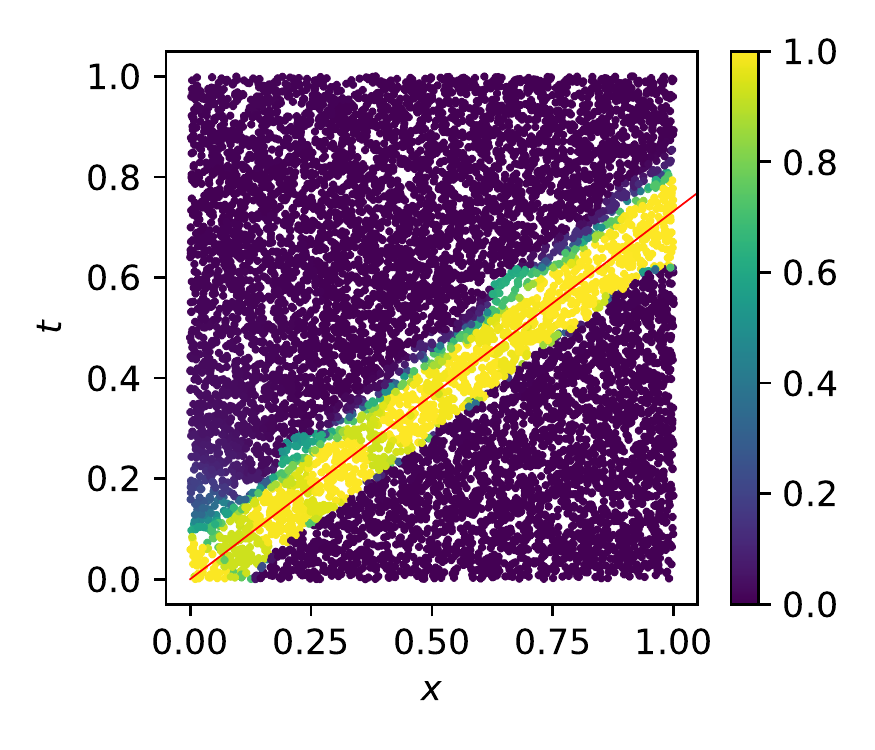}
        \caption{Map}
        \label{fig:av_residual_map}
    \end{subfigure}
    \begin{subfigure}{0.4\textwidth}
        \includegraphics[width=\linewidth]{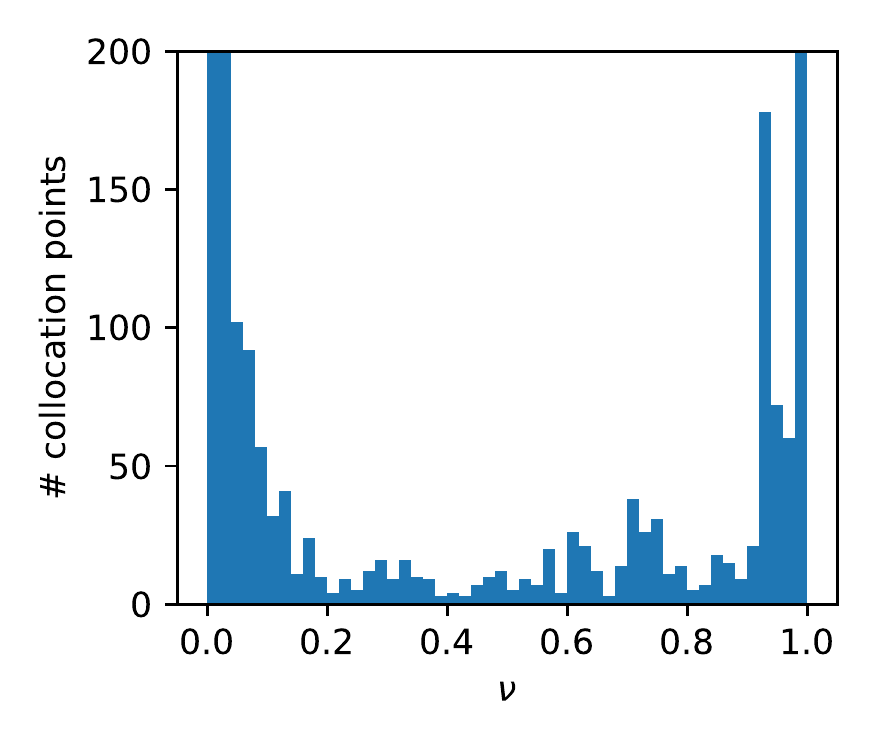}
        \caption{Histogram}
        \label{fig:av_residual_histogram}
    \end{subfigure}
    \caption{(a) Residual-based artificial viscosity map obtained from a trained PINN using fixed unique artificial viscosity value. The red diagonal line represents the shock front path in a hypothetical problem. (b) Residual-based artificial viscosity map histogram.}
    \label{fig:av_residual}
\end{figure}

As an illustration of the proposed method, Figure \ref{fig:av_residual_map} displays the viscosity map calculated by (\ref{eq:av_residual}). It is possible to observe a band of high values around the analytically calculated shock front position (red line). The width of this band is controlled by the number of neighbors points used in the calculations of the $\max_{loc}$ and $\min_{loc}$ operators. The map displayed here was built using 100 neighbors. Figure \ref{fig:av_residual_histogram}(b) depicts the histogram of the map values over the entire domain, confirming that most of the values are concentrated around the extreme values of no viscosity and maximum viscosity. 

\section{Experimental Results}
\label{sec:results}

This section presents experimental results obtained with the three proposed AV methods.
As a warm-up, we first present results obtained for the standard inviscid Burgers equation benchmark. Then we consider the Buckley-Leverett problem with several flux function types when we also compare our results with those in \citep{fuks_limitations_2020}.

\subsection{Inviscid Burgers Equation}

The one-dimensional inviscid Burgers equation with sinusoidal initial condition and Dirichlet boundary conditions has become a standard benchmark in PINN research. It reads:
\begin{subequations}
    \label{eq:burgers_eq_system}
    \begin{align} 
        &\pdv{u}{t} + u\pdv{u}{x} = 0, \quad x \in [-1,1], t \in [0,1] \label{eq:burgers_pde}\\
        &u(x,t=0) = -\sin{(\pi x)}  \label{eq:burgers_ic} \\
        &u(x=-1,t) = u(x=1,t) = 0  \label{eq:burgers_bc}.
    \end{align}
\end{subequations}

\begin{figure}[!ht]
     \centering
     \includegraphics[width=0.75\linewidth]{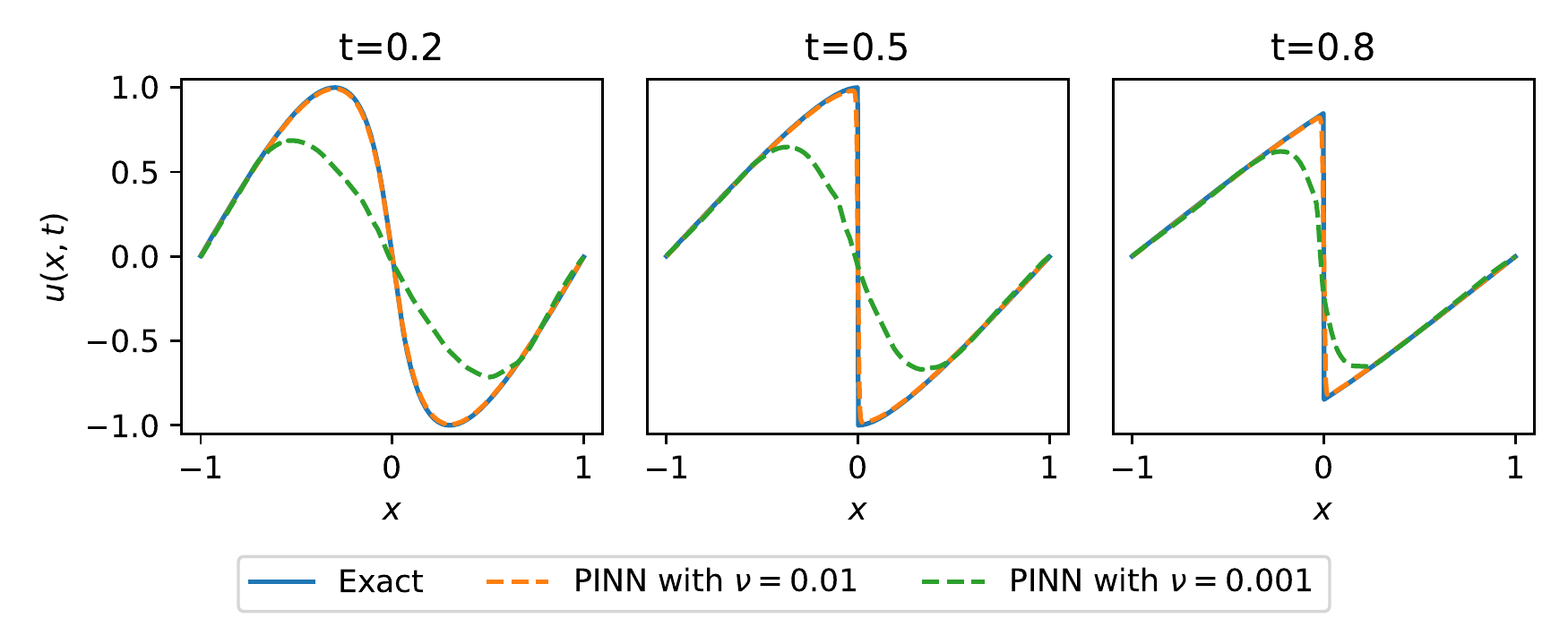}
     \caption{Burgers equation with baseline PINN algorithm. The PINN approximates the solution well with $\nu = 0.01$, but not with $\nu = 0.001$.}
     \label{fig:burgers_raissi}
\end{figure}

The solution of this nonlinear PDE develops a shock at $x=0$, and most numerical methods have problems in capturing this discontinuity. Adding a diffusion term to this equation can mitigate this problem:
\begin{align} 
    \pdv{u}{t} + u\pdv{u}{x} = \frac{\nu}{\pi} \pdv[2]{u}{x}\,,
\label{eq:viscous_Burgers}
\end{align}
where $\nu > 0$ is the viscosity coefficient. 
As was observed by \cite{raissi_physics-informed_2019} and others, the basic PINN algorithm can solve the viscous Burgers PDE in (\ref{eq:viscous_Burgers}) with $\nu = 0.01$. However, if $\nu$ is substantially below $0.01$, the PDE approaches the inviscid one, and the baseline PINN cannot converge to the correct solution, as can be seen in Figure \ref{fig:burgers_raissi}.

Below we discuss results obtained by applying the three proposed adaptive AV methods to the inviscid Burgers problem. The parametric AV map was built using the knowledge that a shock emerges at $x = 0$. This leads to two parameters: the time when the shock forms $t_{shock}$ and the artificial viscosity bandwidth  $w_\nu$, where the values of the maps have a Gaussian distribution $\mathcal{N}(0, w_\nu)$. Here, we set $w_\nu = 0.1$ but learn $t_{shock}$ using the neural network.

Table \ref{table:burgers_error} compares the results with those obtained by the baseline PINN with nonadaptive global viscosity coefficient $\nu = 0.01$. The relative $L2$ and $L1$ errors are computed against a high-fidelity solution of the inviscid Burgers problem provided by the Clawpack-5 finite-volume software\footnote{https://www.clawpack.org/gallery/pyclaw/gallery/burgers\_1d.html}. The results are based on 20 independent runs.

\begin{table}[!ht]
    \centering
    \begin{tabular}{ c|c|c|c|c|c|c } 
        \hline
            \multirow{2}{*}{PINN Method} & \multicolumn{3}{c|}{L2 Error ($\%$)}& \multicolumn{3}{c}{L1 Error ($\%$)}  \\
        \cline{2-7}
            & Mean & Std Dev & Median & Mean & Std Dev & Median\\
        \hline
            baseline PINN, $\nu = 0.01$ & $6.10$ & $0.03$ & $6.09$ & $1.30$ & $0.02$ & $1.30$ \\
            learnable global AV         & $5.73$ & $0.43$ & $5.63$ & $1.21$ & $0.13$ & $1.20$ \\
            parametric AV map           & $5.14$ & $0.43$ & $5.07$ & $0.80$ & $0.10$ & $0.78$ \\
            residual-based AV map       & $7.22$ & $0.60$ & $7.29$ & $1.34$ & $0.20$ & $1.34$ \\ 
         \hline
    \end{tabular}
    \caption{Inviscid Burgers PDE results.}
    \label{table:burgers_error}
\end{table}

The viscosity value learned by the PINN with the global AV method was $0.0088 \pm 0.0010$, which is smaller than the 0.01 value in the nonadaptive method. The value obtained for the learnable maximum viscosity parameter $\nu_{max}$ used in the parametric AV map was $0.0078 \pm 0.0010$, while the learned value for $t_{shock}$ was $0.223 \pm 0.032$. This means that the maximum viscosity applied is below the 0.01 value in the nonadaptive method. The learned value for the maximum viscosity $\nu_{max}$ for the residual-based map was $0.0141 \pm 0.0023$.  
 
\begin{figure}[!ht]
    \centering
    \includegraphics[width=0.75\linewidth]{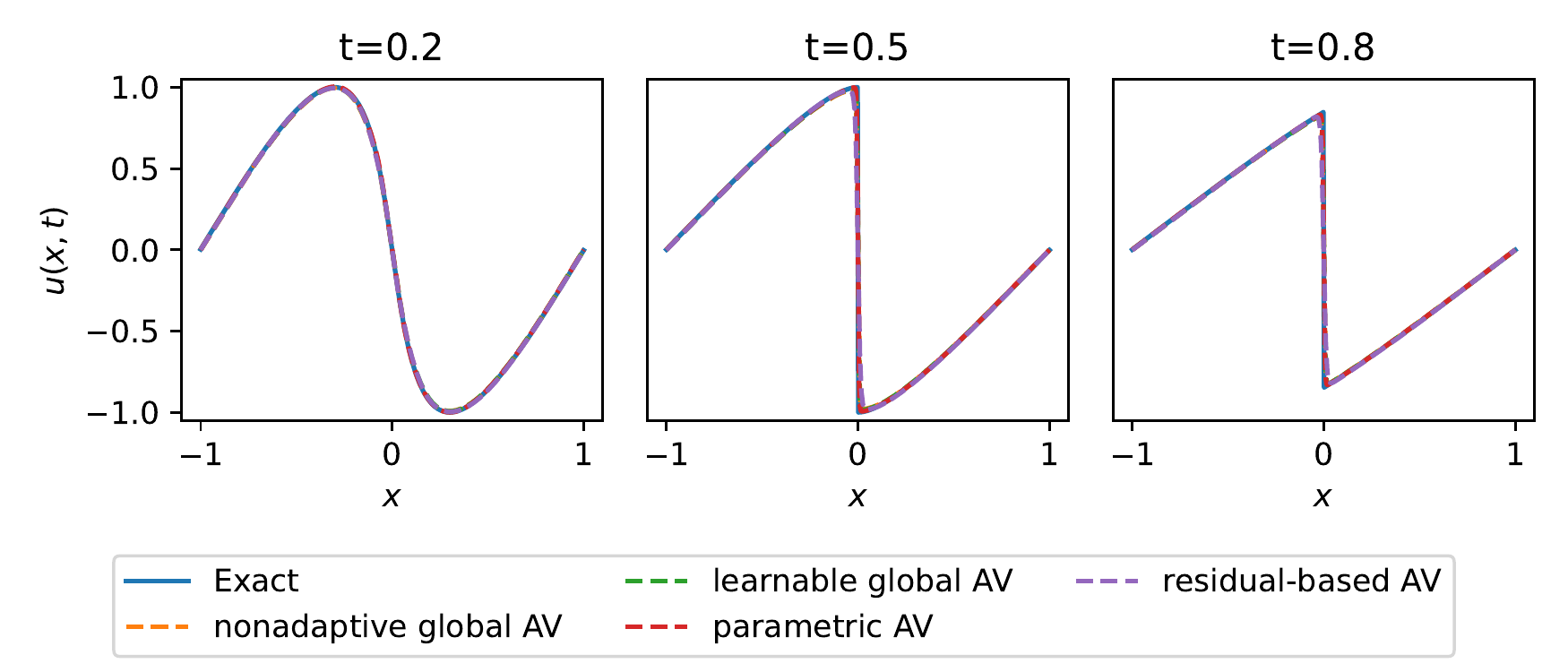}
    \caption{Solution of the inviscid Burgers PDE.}
    \label{fig:burgers_compare}
\end{figure}

Figure \ref{fig:burgers_compare} displays plots of the solutions at different times for the run with L2 error closest to the median error for each method. The results look similar across all methods (but the proposed AV methods do not need to pre-specify the artificial viscosity).

\begin{figure}[!ht]
    \centering
    \begin{subfigure}{0.4\textwidth}
        \includegraphics[width=\linewidth]{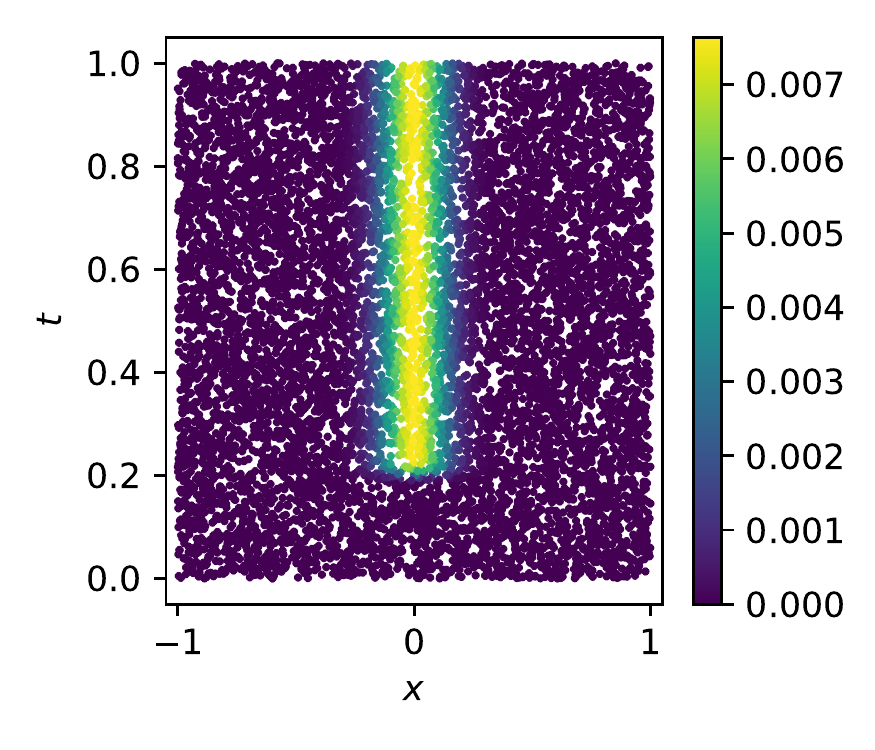}
        \caption{Map}
        \label{fig:burgers_param_map}
    \end{subfigure}
    \begin{subfigure}{0.35\textwidth}
        \includegraphics[width=\linewidth]{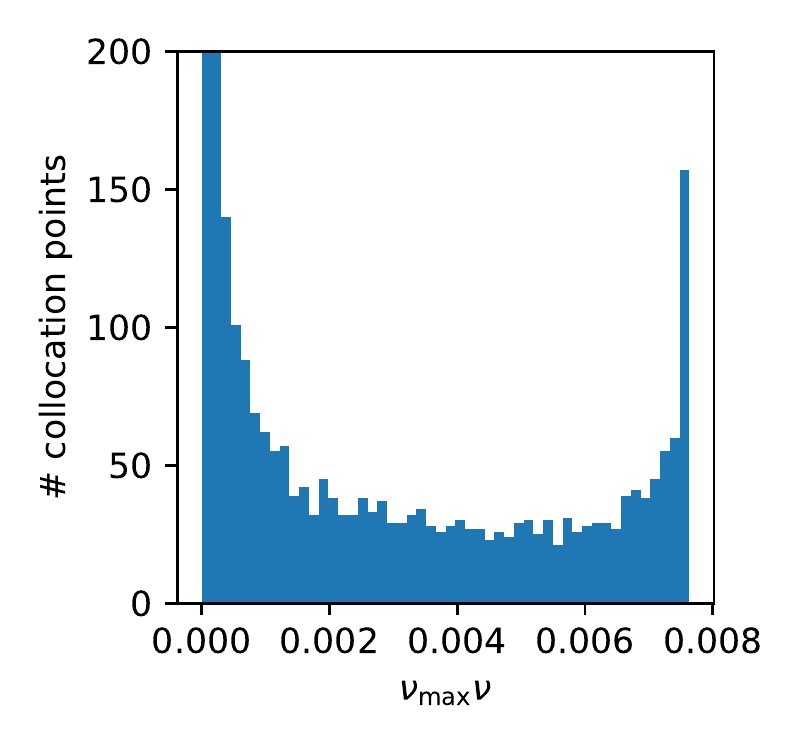}
        \label{fig:burgers_param_histogram}
        \caption{Histogram}
    \end{subfigure}
    \caption{Parametric artificial viscosity map for the inviscid Burgers equation. (a) Map values. (b) Histogram of values.}
\end{figure}

\begin{figure}[!ht]
    \centering
    \begin{subfigure}{0.4\textwidth}
        \includegraphics[width=\linewidth]{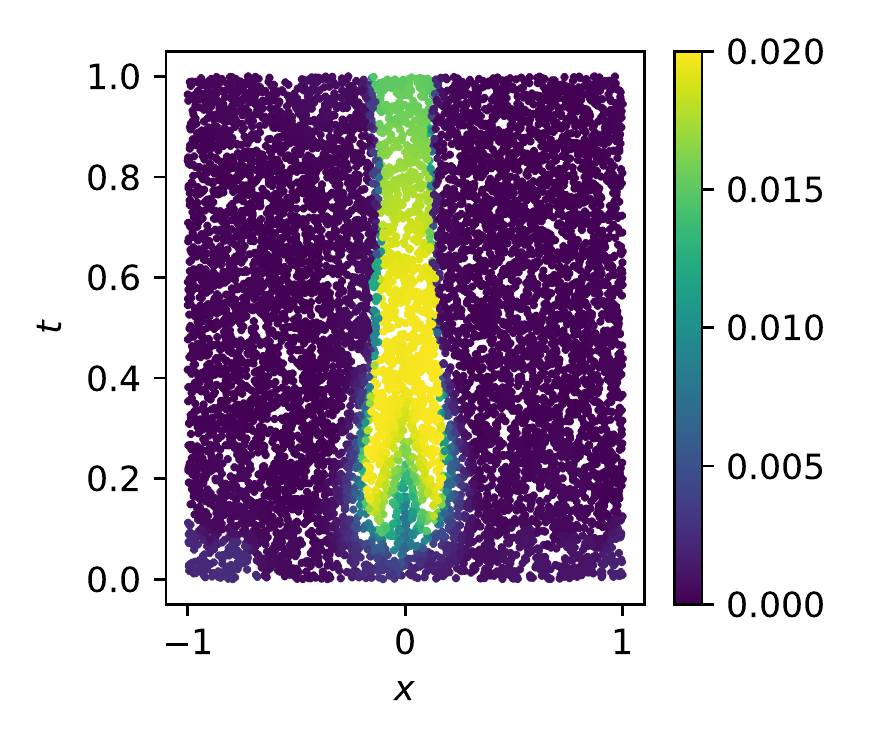}
        \caption{Map}
        \label{fig:burgers_viscosity_map}
    \end{subfigure}
    \begin{subfigure}{0.35\textwidth}
        \includegraphics[width=\linewidth]{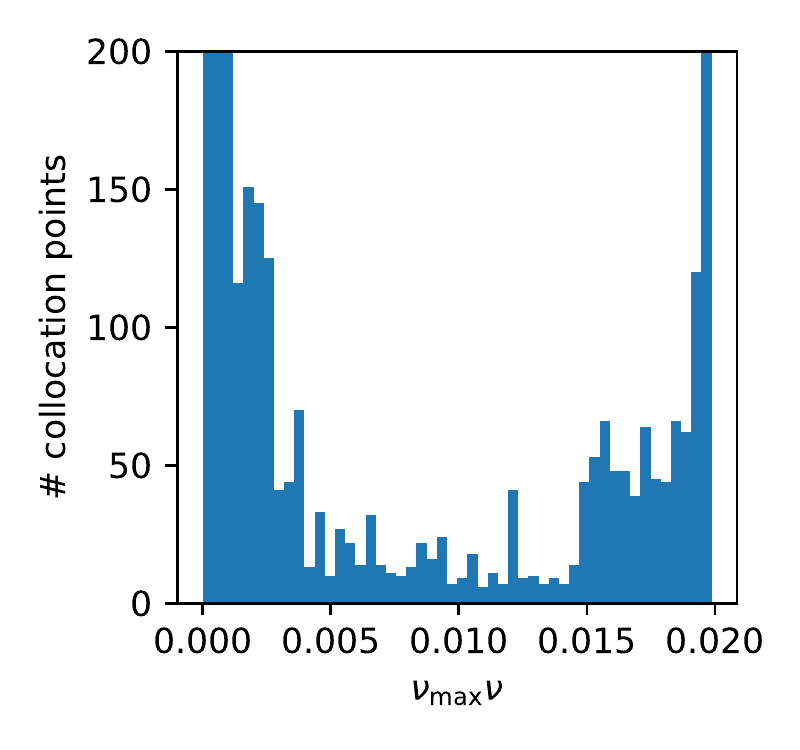}
        \caption{Histogram}
        \label{fig:burgers_viscosity_histogram}
    \end{subfigure}
    \caption{Residual-based artificial viscosity map for the inviscid Burgers equation. (a) Map values. (b) Histogram of values.}
    \label{fig:burgers_residual}
\end{figure}

We can see in Table \ref{table:burgers_error} that the parametric AV map produces the best result both in L1 and L2 error, while the residual-based AV map does not perform as well. To examine this closer, Figures \ref{fig:burgers_param_map} and \ref{fig:burgers_viscosity_map} display the parametric and residual-based AV maps learned during the training procedure for the same run displayed in Figure \ref{fig:burgers_compare}. The values displayed in these maps are already multiplied by the learned value of $\nu_{\max}$ in each case, so they represent the actual viscosity applied to the solution domain. We can see that in both cases, the learned viscosity is only significantly higher than zero in areas close to the shock; however, the parametric map is better localized than the residual-based one. In addition, we can see that the parametric map applies a much smaller amount of viscosity than the residual-based one. (In fact, the parametric AV map produces the smallest values of artificial viscosity among all the methods.) Notice, however, that the residual-based AV map learns the position of the shock without the need for any prior knowledge about the structure of the solution.

\subsection{The Buckley-Leverett problem} 
\label{sec:buckley_leverett}
The Buckley-Leverett problem \citep{buckley_mechanism_1942} is the mathematical model of the displacement by water of immiscible and incompressible oil in porous media. Mass balance equations and Darcy's law for each phase (oil and water) are used and after some manipulation, the one-dimensional problem can be written as:
\begin{subequations}
    \label{eq:bl_eq_system}
    \begin{align} 
        &\pdv{S_w}{t} + \pdv{f_w(S_w)}{x} = 0, \quad x \in [0,1], t \in [0,1] \label{eq:bl_pde}\\
        &S_w(x,t=0) = 0  \label{eq:bl_ic} \\
        &S_w(x=0,t) = 1  \label{eq:bl_bc},
    \end{align}
\end{subequations}
where $S_w$ is the water saturation and $f_w$ is the fractional flow function. The latter is related to the properties of the rock-fluid interaction; its values are usually obtained from fluid viscosity and relative permeability curves. Here, we are going to use simplified fractional flow relations, including concave, convex, and non-convex cases.

\begin{figure}[!ht]
    \begin{subfigure}{0.31\textwidth}
        \includegraphics[width=\linewidth]{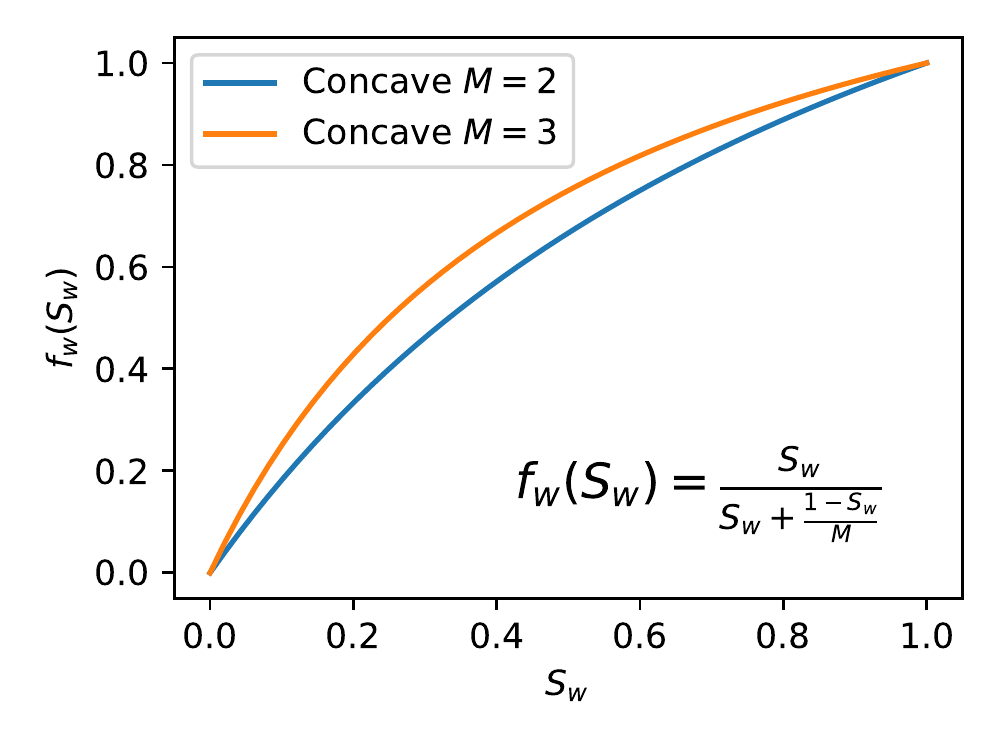}
        \caption{Concave}
        \label{fig:ff_concave}
    \end{subfigure}
    \hspace*{\fill}
    \begin{subfigure}{0.31\textwidth}
        \includegraphics[width=\linewidth]{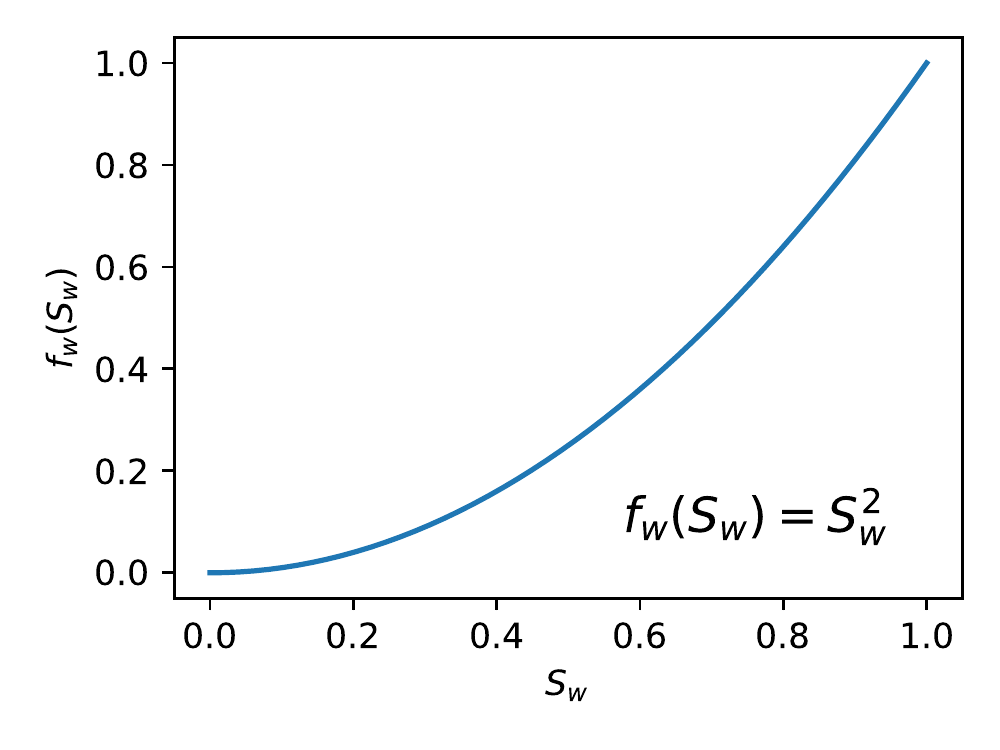}
        \caption{Convex}
        \label{fig:ff_convex}
    \end{subfigure}
    \hspace*{\fill}
    \begin{subfigure}{0.31\textwidth}
        \includegraphics[width=\linewidth]{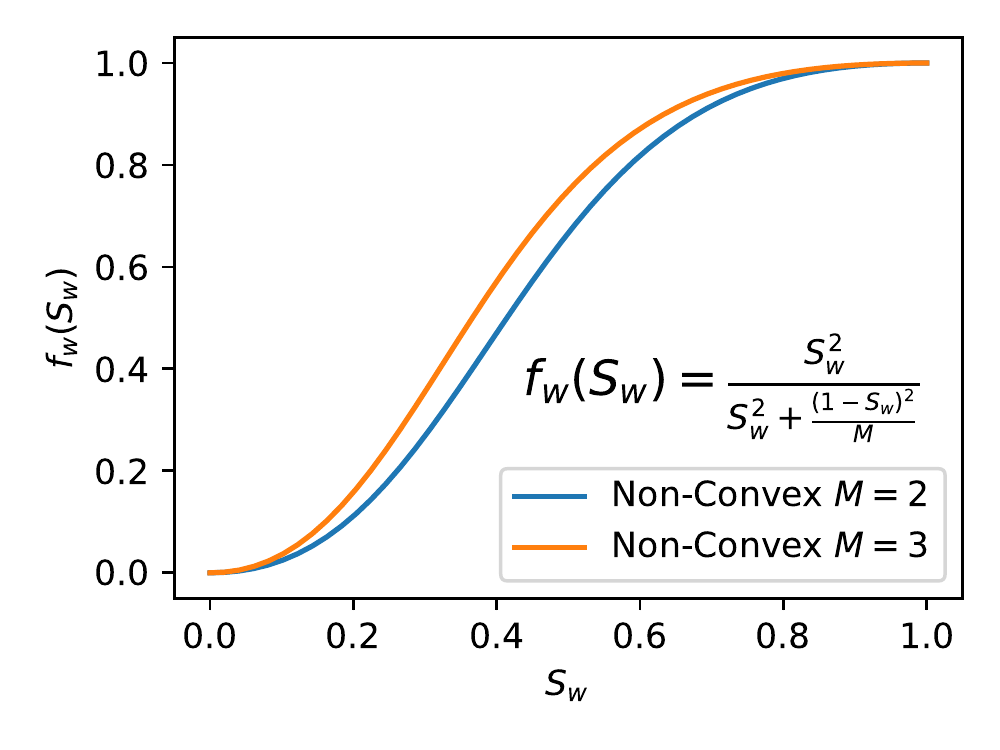}
        \caption{Non-Convex}
        \label{fig:ff_nonconvex}
    \end{subfigure}
    \caption{Fractional flow function types.}
    \label{fig:ff_types}
\end{figure}

\begin{figure}[!ht]
    \begin{subfigure}{0.31\textwidth}
        \includegraphics[width=\linewidth]{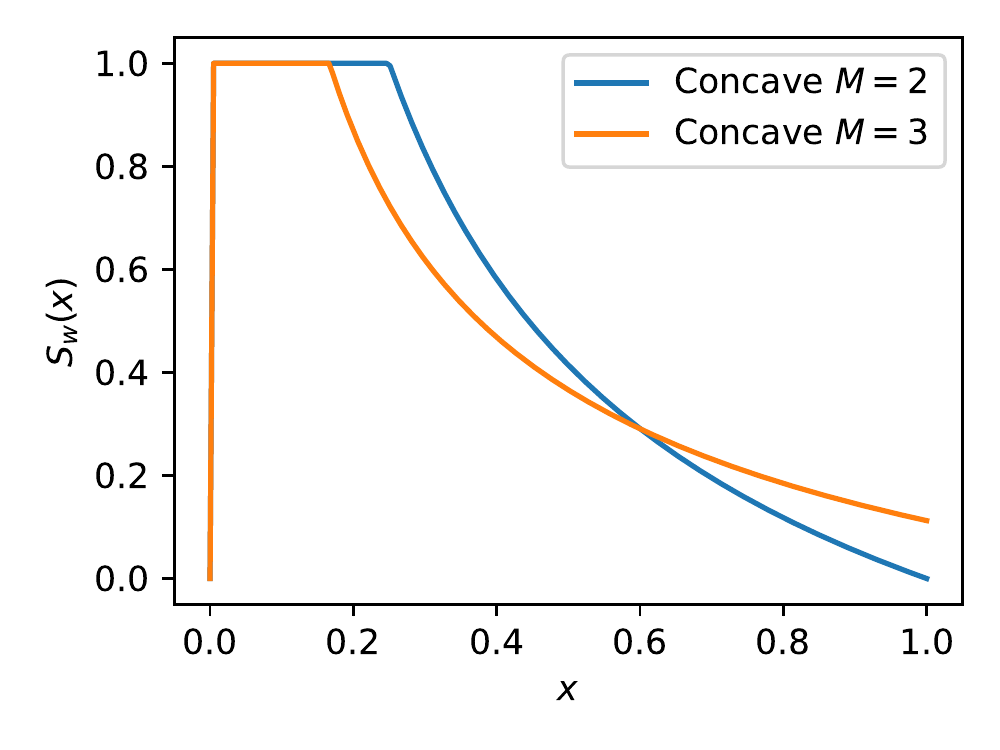}
        \caption{Concave}
        \label{fig:ff_concave_output}
    \end{subfigure}
    \hspace*{\fill}
    \begin{subfigure}{0.31\textwidth}
        \includegraphics[width=\linewidth]{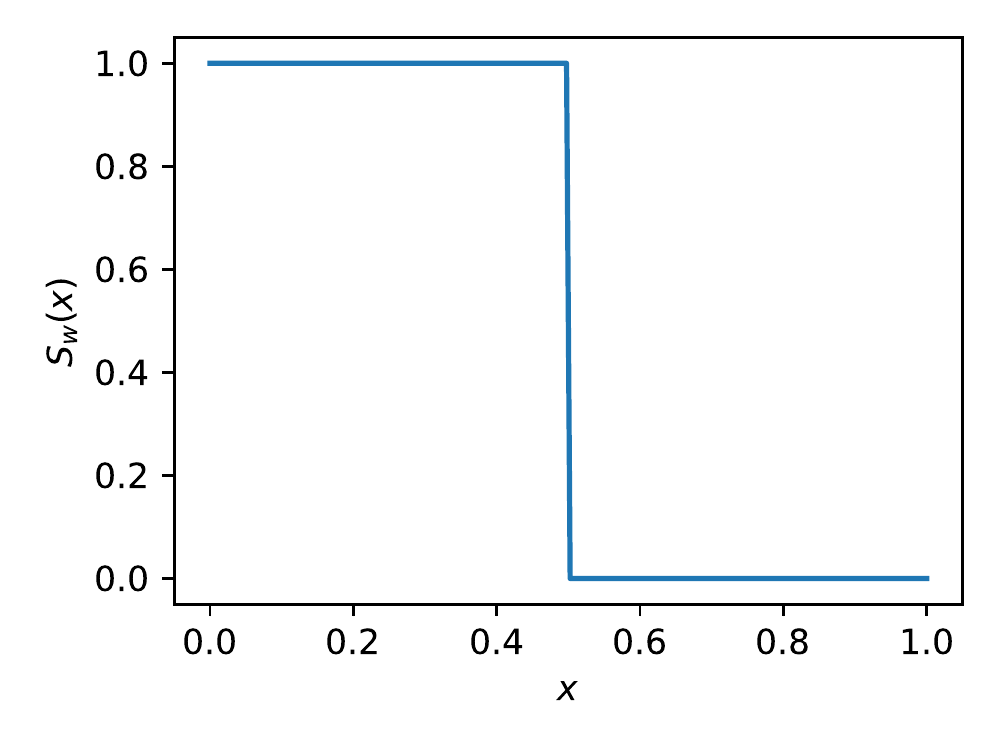}
        \caption{Convex}
        \label{fig:ff_convex_output}
    \end{subfigure}
    \hspace*{\fill}
    \begin{subfigure}{0.31\textwidth}
        \includegraphics[width=\linewidth]{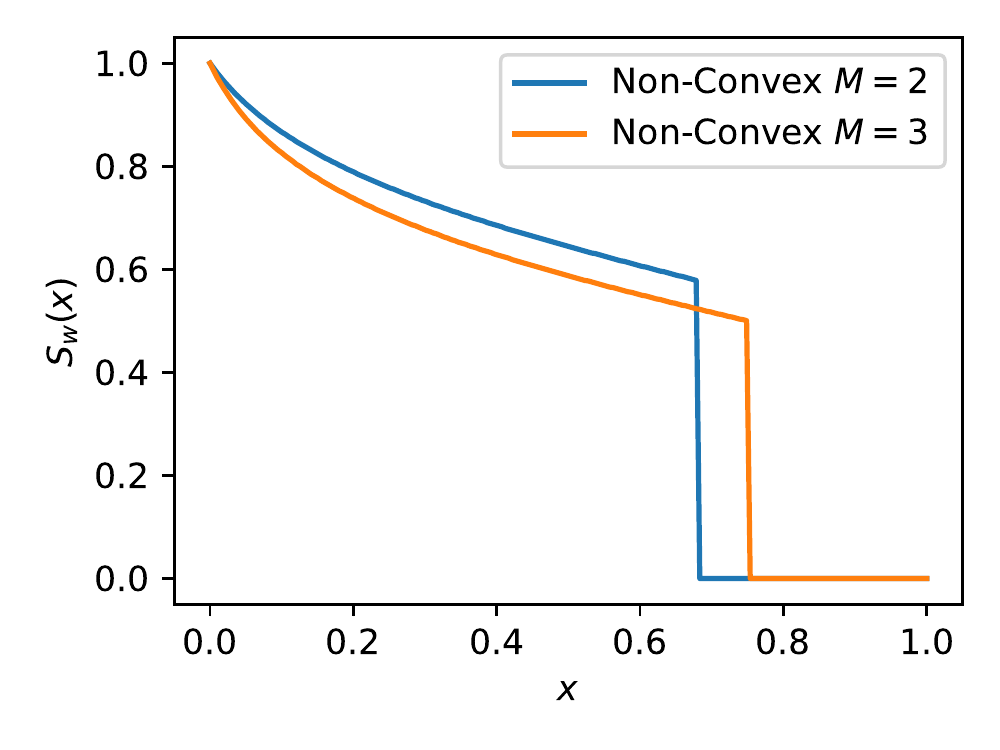}
        \caption{Non-Convex}
        \label{fig:ff_nonconvex_output}
    \end{subfigure}
    \caption{Exact solutions at $t=0.5$ for the fractional flow functions in the previous figure.}
    \label{fig:ff_types_output}
\end{figure}

The concave fractional flow function is defined as:
\begin{equation}
    f_w(S_w) = \frac{S_w}{S_w + \frac{(1-S_w)}{M}}\,, \label{eq:bl_concave_ff}
\end{equation}
which is displayed in Figure \ref{fig:ff_concave}. The parameter $M=\mu_o/\mu_w$ represents the viscosity ratio between the fluid originally in the porous media (oil) and the one used as a displacement fluid (water). This fractional flow function produces a solution that contains a single rarefaction wave, as seen in Figure \ref{fig:ff_concave_output}.

The convex fractional flow function is simple:
\begin{equation}
    f_w(S_w) = S_w^2\,, \label{eq:bl_convex_ff}
\end{equation}
which is displayed in Figure \ref{fig:ff_convex}).  This fractional flow function leads to the presence of a single shock front in the solution, as seen in Figure \ref{fig:ff_convex_output}.

The most interesting case is that of a non-convex fractional flow function, given by: 
\begin{equation}
    f_w(S_w) = \frac{S_w^2}{S_w^2+\frac{(1-S_w)^2}{M}}\,, \label{eq:bl_nonconvex_ff}
\end{equation}
which can be seen in Figure \ref{fig:ff_nonconvex}. The
parameter $M$ is as in the previous case. Figure \ref{fig:ff_nonconvex_output} shows that this  fractional flow function causes the solution to develop simultaneous rarefaction and shock waves. 

\cite{fuks_limitations_2020} showed that PINN could be used to solve the Buckley-Leverett problem with a concave fractional flow function with no artificial viscosity. On the other hand, they also showed that adding artificial viscosity was needed to solve the convex and non-convex fractional flow functions. The main issue with applying PINN to solve the Buckley-Leverett problem, and generally hyperbolic PDEs, is the presence of discontinuities.

We employ three metrics to judge the quality of each proposed model. The L2 error is defined as:
\begin{equation}
    \textrm{L2 Error} = \frac{\norm{S_{w,\text{exact}} - S_w}_2}{\norm{S_{w,\text{exact}}}_2}\,.
\end{equation}
The L1 error is defined similarly, by replacing the L2 norm with the L1 norm. In the cases where a shock is present (convex and non-convex flows), we introduce the {\em front error}, which compares the position of the shock front determined by the PINN to the exact position.
The position of the shock front found by the PINN is estimated as
\begin{equation}
    x_{\text{front}} = \argmin_x{\pdv{}{x}S_w(x,t,\vb*{w})}, 
\end{equation}
where $\pdv{S_w}{x}$ can be obtained by automatic differentiation using the trained neural network. The exact shock front position $x_{\text{front,exact}}$ is easily calculated since the front velocity is constant and known. The front error is then defined as
\begin{equation}
    \textrm{front error} = \frac{\norm{x_{\text{front,exact}} - x_{\text{front}}}_2}{\norm{x_{\text{front,exact}}}_2}
\end{equation}

Table \ref{table:bl_all_methods_errors} displays the results obtained with the nonadaptive global AV method in \cite{
fuks_limitations_2020} and the three proposed adaptive AV methods. In all experiments, each training run was repeated 20 times with different random number generator seeds, and the mean, standard deviation, and median of the three error metrics for each method are reported. We discuss below the results obtained by each of the three proposed methods.

\begin{table}[!ht]
    \small
    \centering
    \setlength\tabcolsep{3.0pt} 
    \begin{tabular}{ c|c|c|c|c|c|c|c|c|c|c} 
        \hline
            \multirow{2}{*}{Flux Type} &
            \multirow{2}{*}{PINN Method} & \multicolumn{3}{c|}{L2 Error (\%)} & \multicolumn{3}{c|}{L1 Error (\%)}& \multicolumn{3}{c}{Front Error (\%)} \\
        \cline{3-11}
            & & Mean & Std Dev & Median & Mean & Std Dev & Median & Mean & Std Dev & Median\\
        \hline
            \multirow{4}{*}{non-Convex $M=0.5$} &
            nonadaptive global AV  & $6.55$ & $0.97$ & $6.76$ & $1.20$ & $0.13$ & $1.19$ & $1.20$ & $0.27$ & $1.25$ \\
            &learnable global AV   & $6.72$ & $1.23$ & $6.60$ & $1.16$ & $0.31$ & $1.11$ & $1.21$ & $0.34$ & $1.16$ \\
            &parametric AV map     & $6.93$ & $1.04$ & $7.13$ & $1.40$ & $0.31$ & $1.45$ & $1.29$ & $0.32$ & $1.29$ \\
            &residual-based AV map & $8.70$ & $1.38$ & $8.78$ & $2.23$ & $0.51$ & $2.20$ & $1.97$ & $0.51$ & $1.99$ \\
        \hline
            \multirow{4}{*}{non-Convex $M=1$} &
            nonadaptive global AV  & $7.23$ & $0.84$ & $7.31$ & $1.46$ & $0.19$ & $1.43$ & $1.66$ & $0.30$ & $1.66$ \\
            &learnable global AV   & $7.63$ & $1.06$ & $7.44$ & $1.45$ & $0.31$ & $1.37$ & $1.71$ & $0.46$ & $1.59$ \\
            &parametric AV map     & $6.42$ & $1.66$ & $6.53$ & $1.23$ & $0.37$ & $1.22$ & $1.36$ & $0.57$ & $1.35$ \\
            &residual-based AV map & $10.04$ & $1.27$ & $10.33$ & $2.48$ & $0.38$ & $2.51$ & $3.45$ & $0.78$ & $3.53$ \\
        \hline
            \multirow{4}{*}{non-Convex $M=2$} & 
            nonadaptive global AV  & $7.05$ & $0.97$ & $7.22$ & $1.80$ & $0.30$ & $1.76$ & $2.48$ & $0.58$ & $2.53$ \\
            &learnable global AV   & $6.88$ & $0.68$ & $6.96$ & $1.62$ & $0.22$ & $1.65$ & $2.25$ & $0.39$ & $2.29$ \\
            &parametric AV map     & $5.94$ & $1.28$ & $5.97$ & $1.34$ & $0.34$ & $1.26$ & $1.80$ & $0.67$ & $1.67$ \\
            &residual-based AV map & $8.33$ & $0.57$ & $8.22$ & $2.14$ & $0.21$ & $2.12$ & $3.55$ & $0.43$ & $3.48$ \\
        \hline
            \multirow{4}{*}{non-Convex $M=5$} &
            nonadaptive global AV  & $7.22$ & $1.36$ & $7.14$ & $2.86$ & $0.55$ & $2.76$ & $3.91$ & $1.33$ & $3.78$ \\
            &learnable global AV   & $6.47$ & $1.13$ & $6.45$ & $2.45$ & $0.41$ & $2.44$ & $3.14$ & $0.98$ & $3.06$ \\
            &parametric AV map     & $6.56$ & $0.99$ & $6.31$ & $2.00$ & $0.34$ & $1.91$ & $3.34$ & $0.90$ & $3.06$ \\
            &residual-based AV map & $7.28$ & $0.72$ & $7.49$ & $2.75$ & $0.34$ & $2.77$ & $4.31$ & $0.75$ & $4.48$ \\
        \hline
            \multirow{4}{*}{non-Convex $M=10$} &
            nonadaptive global AV  & $8.75$ & $1.83$ & $8.73$ & $4.66$ & $1.01$ & $4.64$ & $17.45$ & $18.06$ & $9.08$ \\
            &learnable global AV   & $7.52$ & $1.24$ & $7.32$ & $3.85$ & $0.65$ & $3.78$ & $7.77$ & $9.58$ & $5.75$ \\
            &parametric AV map     & $7.66$ & $2.37$ & $7.12$ & $3.35$ & $1.17$ & $3.08$ & $15.34$ & $23.79$ & $6.80$ \\
            &residual-based AV map & $7.19$ & $1.57$ & $6.75$ & $3.78$ & $0.86$ & $3.51$ & $8.38$ & $13.98$ & $4.98$ \\
        \hline
            \multirow{4}{*}{Convex} &
            nonadaptive global AV  & $4.50$ & $0.25$ & $4.42$ & $0.79$ & $0.03$ & $0.78$ & $0.27$ & $0.15$ & $0.20$ \\
            &learnable global AV   & $3.05$ & $0.44$ & $2.91$ & $0.32$ & $0.09$ & $0.30$ & $0.16$ & $0.04$ & $0.15$ \\
            &parametric AV map     & $3.33$ & $1.09$ & $3.00$ & $0.36$ & $0.21$ & $0.30$ & $0.22$ & $0.16$ & $0.20$ \\
            &residual-based AV map & $3.34$ & $0.58$ & $3.07$ & $0.39$ & $0.11$ & $0.34$ & $0.19$ & $0.11$ & $0.15$ \\
         \hline
            \multirow{4}{*}{Concave M = 2.0} &
            nonadaptive global AV  & $6.95$ & $0.01$ & $6.95$ & $0.60$ & $0.08$ & $0.62$ & $-$ & $-$ & $-$ \\
            &learnable global AV   & $6.96$ & $0.01$ & $6.96$ & $0.63$ & $0.05$ & $0.63$ & $-$ & $-$ & $-$ \\ 
            &parametric AV map     & $6.95$ & $0.01$ & $6.95$ & $0.61$ & $0.08$ & $0.60$ & $-$ & $-$ & $-$ \\
            &residual-based AV map & $6.95$ & $0.01$ & $6.95$ & $0.61$ & $0.08$ & $0.58$ & $-$ & $-$ & $-$ \\
         \hline
    \end{tabular}
    \caption{Prediction error statistics for the nonadaptive global AV method in \cite{fuks_limitations_2020} and the three proposed adaptive AV methods applied to the Buckley-Leverett problem.}
    \label{table:bl_all_methods_errors}
\end{table}

\subsection{PINN with Learnable Global Artificial Viscosity}
\label{sec:results_pinn_learnable_av}

\begin{figure}[!ht]
    \centering
    \includegraphics[width=\linewidth]{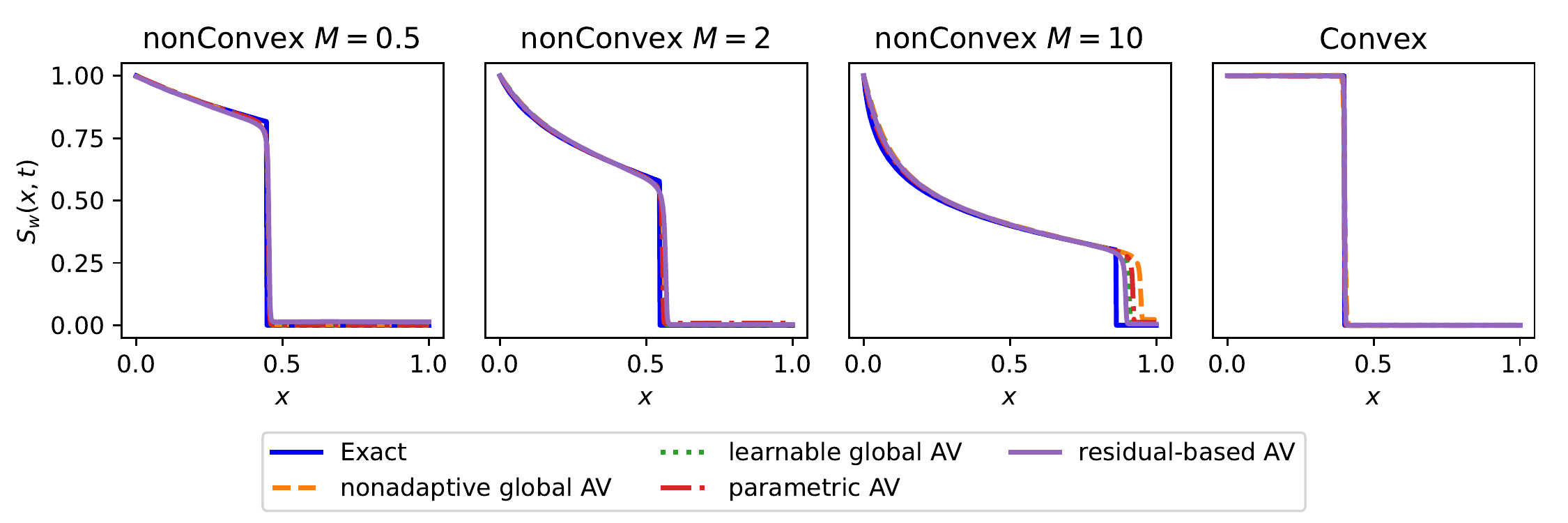}
    \caption{Predictions for the nonadaptive global AV method in \cite{fuks_limitations_2020} and the three proposed adaptive AV methods applied to the Buckley-Leverett problem at $t=0.4$}
    \label{fig:bl_pinn_all_methods}
\end{figure}

Here we contrast the learnable and nonadaptive global AV methods. Table \ref{table:bl_all_methods_errors} reveals that learning the AV value leads to better L1 errors than using a nonadaptive AV value in all non-convex and convex cases. It produces better L2 errors and front errors than the nonadaptive global AV method in the non-convex case with larger values of $M$, and in the convex case. The concave case has no shock fronts and the two methods produce almost identical errors. Table \ref{table:bl_learnable_param_learned} displays the statistics of the artificial viscosity learned for each flux type. We can observe that the values learned vary considerably among the different flux types. This is unsurprising since each different flux type presents a different shock configuration. In the concave flow case, the method learned a viscosity close to zero, which is entirely consistent with the fact that we do not need a diffusion term to use PINN to solve the problem. Notice that the learned AV values are in all cases smaller than the nonadaptive value used in \cite{fuks_limitations_2020}. Figure \ref{fig:bl_pinn_all_methods} displays sample values of the solutions obtained with the two methods for a few different flow cases.

\begin{table}[!ht]
    \centering
    \begin{tabular}{ c|c|c } 
        \hline
            \multirow{2}{*}{Flux Type} & \multicolumn{2}{c}{$\nu \quad (10^{-3})$} \\
        \cline{2-3}
            & Mean & Std Dev \\
        \hline
            non-Convex $M = 0.5$ & $1.963$ & $0.289$ \\
            non-Convex $M = 1$   & $1.882$ & $0.267$ \\
            non-Convex $M = 2$   & $1.889$ & $0.287$ \\
            non-Convex $M = 5$   & $2.217$ & $0.411$  \\
            non-Convex $M = 10$  & $2.376$ & $0.331$  \\
            Convex               & $0.933$ & $0.095$  \\
            Concave $M = 2$      & $0.033$ & $0.027$  \\
         \hline
    \end{tabular}
    \caption{Artificial viscosity estimated by the PINN with learnable global artificial viscosity. For reference, the nonadaptive artificial viscosity value used in \cite{fuks_limitations_2020} is $2.5 \times 10^{-3}$. In the concave case, the reference value is zero.}
    \label{table:bl_learnable_param_learned}
\end{table}

\subsection{PINN with Parametric Artificial Viscosity Map}
\label{sec:results_pinn_param_map}

The PINN with learnable global artificial viscosity method is able to learn an AV value that produces, in most cases, better results than the PINN with a nonadaptive global AV However, this method applies the AV to the whole solution domain. Table \ref{table:bl_all_methods_errors} shows that 
one can successfully learn the parameters to build an AV map that will localize the application of the AV only where is necessary, close to the discontinuities. 
In several of the cases, this produced the best result among all four alternatives. Both L1 and L2 errors for the non-convex (exception $M=0.5$), convex and concave have lower or very close errors when comparing with all methods. The front error for the non-convex $M=1$ and $2$ present significantly lower values when compared with the other methods.
In this application, the parameters to be learned to construct the AV map are the maximum artificial viscosity value and the shock front velocity; we set the artificial viscosity bandwidth to $w_\nu = 0.1$. The parameters were initialized to $\nu_{max} = 0.0$ and $v_{\text{shock}} = 1.0$.  The statistics of the learned variables are presented in Table~\ref{table:bl_parameterized_param_learned}. The values for the maximum viscosity are smaller than the global nonadaptive method, except in the non-convex case with large $M$. As expected, the maximum viscosity for the concave case is close to zero. Notice that the shock front velocity estimation is quite accurate in all cases. 

\begin{figure}[!ht]
    \centering
    \begin{subfigure}{0.4\textwidth}
        \includegraphics[width=\linewidth]{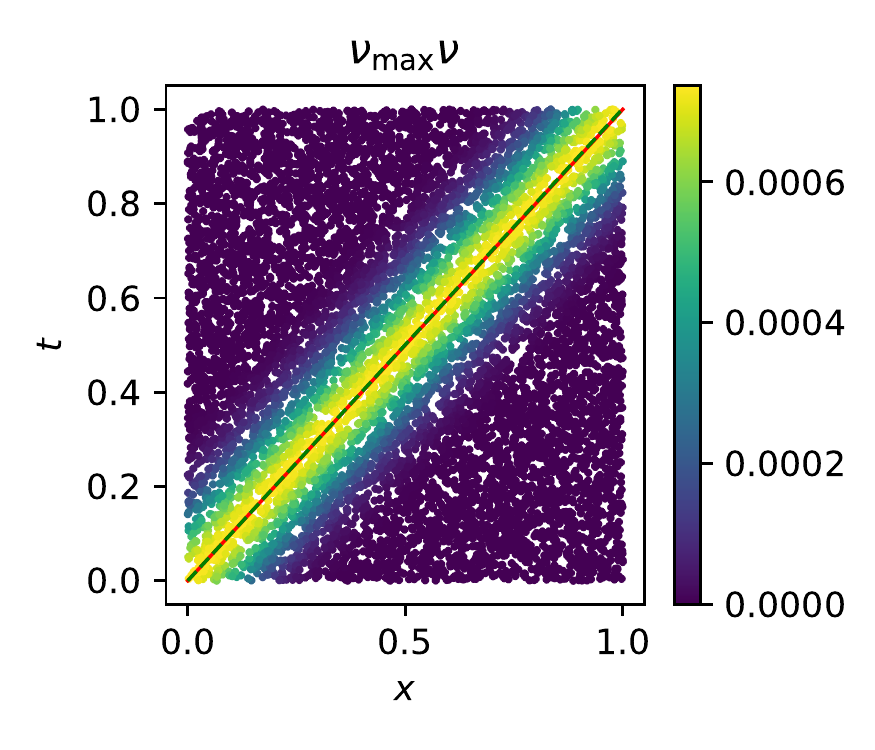}
        \caption{Convex}
        \label{fig:bl_parameterized_visco_map_convex}
    \end{subfigure}
    \begin{subfigure}{0.4\textwidth}
        \includegraphics[width=\linewidth]{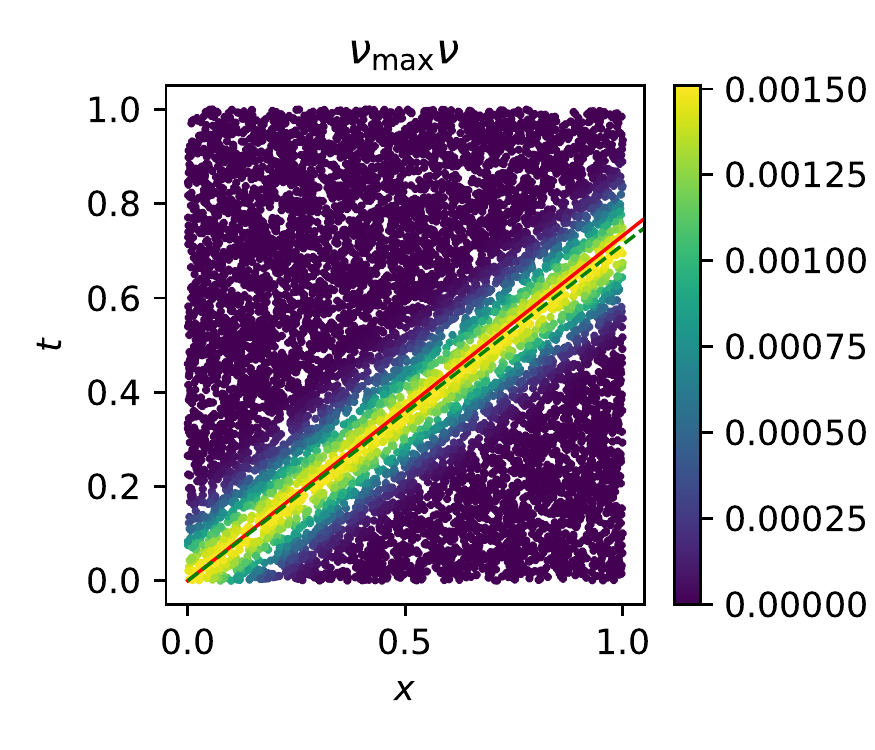}
        \caption{non-Convex $M = 2$}
        \label{fig:bl_parameterized_visco_map_nonconvex}
    \end{subfigure}
    \caption{Artificial Viscosity map produced by the learnable parameters for the Convex and non-Convex $M=2$ flux types. The red line represents the true shock front path with velocity obtained by the analytical solution ($v_{shock} = 1.0$ for convex and $1.366$ for non-Convex with $M=2$). The green dashed line represents the shock path with the learned shock front velocity ($1.0$ for convex and $1.40$ for the non-Convex with $M=2$). The models presented here have L2 error closest to the median for each flux type.}
    \label{fig:bl_parameterized_visco_map}
\end{figure}

Figure \ref{fig:bl_parameterized_visco_map} displays the artificial viscosity map learned for the models that produced L2 error closest to the median of the L2 error distribution. In Figure \ref{fig:bl_parameterized_visco_map_convex}, the Convex fractional flow is present. The shock front velocity for this case is $1.0$ (that generates the red line). It is possible to observe that the band of artificial viscosity covers the evolution of the shock front perfectly. The green dashed line represents the shock front path from the learned front velocity. In Figure \ref{fig:bl_parameterized_visco_map_nonconvex}, the map for the non-Convex case with $M=2$ is presented.

\begin{table}[!ht]
    \centering
    \begin{tabular}{ c|c|c|c|c|c } 
        \hline
            \multirow{2}{*}{Flux Type} & \multicolumn{2}{c|}{$\nu_{max} \quad (10^{-3})$} & \multicolumn{3}{c}{$v_{shock}$} \\
        \cline{2-6}
            & Mean & Std Dev & Mean & Std Dev & Exact\\
        \hline
            non-Convex $M = 0.5$ & $2.020$ & $0.440$ &  $1.137$ & $0.020$ & $1.112$ \\
            non-Convex $M = 1$   & $1.957$ & $0.384$ &  $1.173$ & $0.034$ & $1.207$ \\
            non-Convex $M = 2$   & $1.936$ & $0.376$ &  $1.321$ & $0.052$ & $1.366$ \\
            non-Convex $M = 5$   & $2.022$ & $0.164$ &  $1.775$ & $0.039$ & $1.725$ \\
            non-Convex $M = 10$  & $2.662$ & $1.238$ &  $2.119$ & $0.286$ & $2.158$ \\
            Convex               & $0.842$ & $0.156$ &  $1.000$ & $0.003$ & $1.000$ \\
            Concave $M = 2$      & $0.022$ & $0.027$ &  $-$     & $-$     & $-$ \\
         \hline
    \end{tabular}
    \caption{Parameters estimated by the PINN with parametric artificial viscosity map. For reference, the nonadaptive artificial viscosity value used in \cite{fuks_limitations_2020} is $2.5 \times 10^{-3}$. In the concave case, the reference value is zero.}
    \label{table:bl_parameterized_param_learned}
\end{table}

The results generated by the parametric AV map method are encouraging; however, we need to make some assumptions to build the map. We assumed that only one shock front would happen for the Buckley-Leverett problem and that this shock front would be created right after time zero. In other words, we need to know some information about the solution, use this information to choose the parameters to build the map, and then use PINNs to solve the whole problem by learning these parameters. The results presented in the next section show that a residual-based AV map can overcome this problem, and its application does not rely on any information about the PDE solution.

\subsection{PINN with Residual-based Artificial Viscosity Map}
\label{sec:results_pinn_residual_map}

\begin{figure}[!ht]
    \centering
    \includegraphics[width=\linewidth]{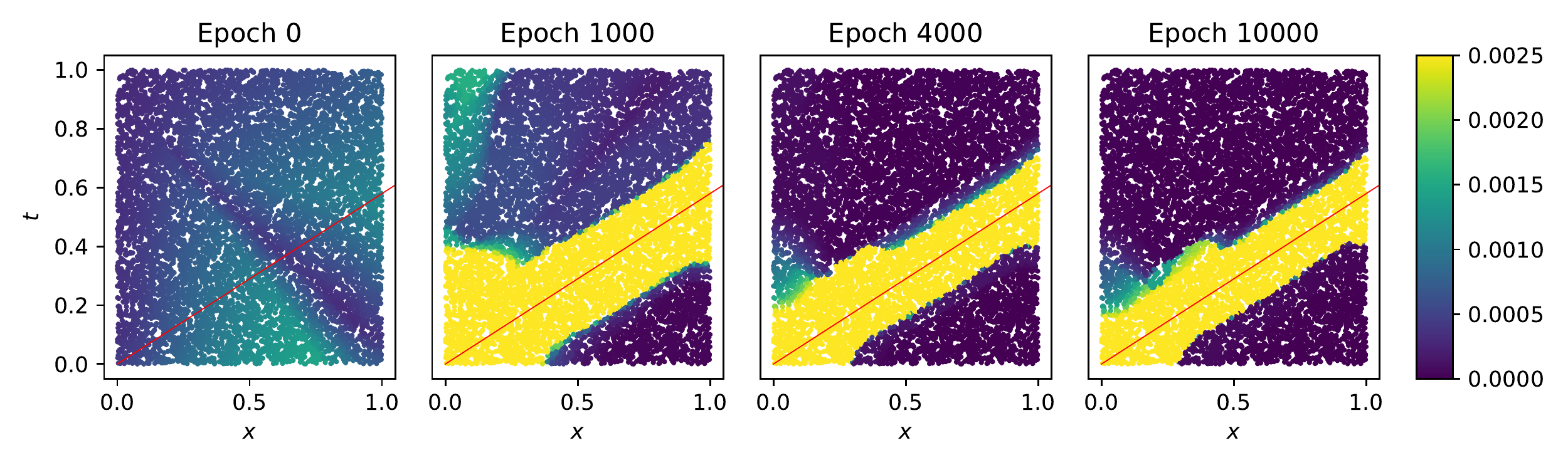}
    \caption{Residual-based AV map evolution during the training process}
    \label{fig:bl_av_map_epochs}
\end{figure}

The application of the PINN with residual-based AV method produced low errors for the non-convex with $M=5$ and $10$ and for the convex cases. For the concave case, like the other methods, the residual-based AV method learned a viscosity map very close to zero. The error statistics for this method applied to all Buckley-Leverett cases are presented in Table \ref{table:bl_all_methods_errors}. The residual-based AV method utilizes an AV map (built from Eq. \ref{eq:av_residual}) that changes along with the PINN training procedure. We observe that the residual-map approach produces the best L2 Error results for the non-convex case at large values of $M$, being less competitive in other cases. Once again, this method does not require any assumptions about the structure of the solution, so a trade-off in accuracy is not unexpected. Figure \ref{fig:bl_av_map_epochs} displays the evolution of the residual map as a function of the training epoch for the non-Convex $M=5$ case. Early in training, it is expected that the method could not localize the AV correctly since the inviscid residual produced by \ref{eq:res_inviscid} is not calibrated. After about 4,000 epochs, the residual presents a higher value close to the discontinuity. See Figure \ref{fig:bl_pinn_all_methods} for sample predictions produced by the residual-based AV map method.

\subsection{Hyperparameter search}
\label{sec:results_hyperparameter_search}

Accurately solving hyperbolic PDE with shocks by means of PINNs requires a judicious choice of the N.N. architecture, optimizer, learning rate, and weights of the penalty terms in the loss functions, among other hyperparameters to be tuned. Each of these parameters plays a different role, and their evolution can often be contradictory to one another in the training phase. We comment below on key hyperparameters and the rationale for choosing them.

The general training  loss function (Eq. \ref{eq:pinn_loss_local}) contains a penalty term $\alpha_{visc}$ to be applied to the artificial viscosity loss term. The choice of this parameter can strongly influence the quality of the final models. Imposing constant values by trial and error could be an option, but we avoid taking this route as this would add another hyperparameter to be tuned in the training procedure. 
Instead, all the results presented here were obtained by setting $\alpha_{visc}$ as a variable in the learning procedure. Its value is initialized with 1.0 and is updated during training. By applying the total loss function  of the negative of gradients' with respect to the $\alpha_{visc}$ value in the Adam update step for gradient ascent, we force this variable to increase during the training procedure. 

Another important hyperparameter to be tuned is the optimization algorithm learning rate. In most cases, the models that produced lower prediction errors were trained with a fixed learning rate of $0.005$ for the neural network weights and biases, and $0.0005$ for the $\nu$, $\nu_{\max}$ and $\alpha_{visc}$ parameters. Attempts to use exponential decaying learning rate schedules or adaptive learning rates did not produce significant improvements. 

Regarding the optimizer's choice, we believe that more work needs to be done to select the best method. Here we employed the general-purpose Adam optimization algorithm employed in general deep learning, which has also been the optimizer of choice in most PINN research at this time. As noted in the text, the proposed methods produced a larger standard deviation than the nonadaptive global AV method in some of the results. We attribute this to the fact that the proposed methods require a more complex optimization problem since they involve learning key parameters, such as AV values and localization parameters, together with the artificial neural network weights. We foresee that a customized optimization scheme would provide more reliable and consistent results using the proposed methods.

\section{Conclusions}
\label{sec:conclusions}

We have proposed in this paper three new methods to add artificial viscosity to train PINNs to solve hyperbolic PDEs with shocks. The methods applications show promising results towards accurately learning both the value and location of the artificial viscosity needed to train the PINNs. All methods proposed here add little extra time to baseline PINN training since they require adding one or two variables to the neural network training procedure.

The application of the learnable global artificial viscosity on all the scenarios for the Buckley-Leverett equation shows that we can obtain similar or lower prediction errors when compared to the nonadaptive choice of a global AV. The same conclusion can be extended to the Inviscid Burger Equation. The parametric AV map method shows that one can localize the AV application on the spatiotemporal domain with some additional knowledge of the solution, providing better results than the one with a learnable global AV. The residual-based method performed reasonably well in most of the cases, particularly in the non-convex case with larger $M$ values. Importantly, all the proposed methods were able to produce sensible results in the Buckley-Leverett problem with a concave fractional flow. In this case, no AV is necessary to train the PINN to solve the problem, and the three methods were capable of learning a global or localized AV close to zero.

We attribute the disappointing performance of the residual-based AV method in some of the settings to limitations in the optimization scheme used to train the N.N. and the necessary additional parameters. Despite that, the residual-based map approach remains a compelling choice because it can localize the AV application and requires no assumptions on the structure of the solution.

\section{Acknowledgments}

Portions of this research were conducted using the advanced computing resources provided by Texas A\&M High-Performance Research Computing. The methods proposed in this research were implemented on a modified version of the TensorDiffEq package \citep{mcclenny_tensordiffeq_2021}. The first author also acknowledges all the support provided by Petroleo Brasileiro S.A.

\bibliographystyle{unsrtnat}
\bibliography{references_zotero, references_extras}  

\end{document}